 \documentclass[final,5p,times,twocolumn]{elsarticle}
\usepackage{amssymb}
\usepackage{setspace}
\usepackage{epstopdf}

\journal{Physics Letters B}

\begin{document}

\begin{frontmatter}



\title{$K^{+}\Lambda$ and $K^+\Sigma^0$ photoproduction with fine center-of-mass energy resolution}
\author[Edinburgh]{T.C.~Jude\fnref{label1}}
\author[Edinburgh]{D.I.~Glazier\fnref{label2}}
\author[Edinburgh]{D.P.~Watts\fnref{label3}}
\author[Mainz]{P.~Aguar-BartolomÂ´e}
\author[Mainz]{L.K.~Akasoy}
\author[Glasgow]{J.R.M.~Annand}
\author[Mainz]{H.J.~Arends}
\author[Kent]{K.~Bantawa}
\author[HISKP]{R.~Beck}
\author[Petersburgh]{V.S.~Bekrenev}
\author[Giessen]{H.~BerghÂ¨auser}
\author[Pavia]{A.~Braghieri}
\author[Edinburgh]{D.~Branford}
\author[Washington]{W.J.~Briscoe}
\author[UCLA]{J.~Brudvik}
\author[Lebedev]{S.~Cherepnya}
\author[Washington]{B.T.~Demissie}
\author[Basel]{M.~Dieterle}
\author[Mainz,Washington]{E.J.~Downie}
\author[Lebedev]{L.V.~Fil'kov}
\author[Giessen]{R.~Gregor}
\author[Mainz]{E.~Heid}
\author[Sackville]{D.~Hornidge}
\author[Basel]{I.~Jaegle}
\author[Mainz]{O.~Jahn}
\author[Mainz,Lebedev]{V.L.~Kashevarov}
\author[Basel]{I.~Keshelashvili}
\author[Moscow]{R.~Kondratiev}
 \author[Zagreb]{M.~Korolija}
 \author[Petersburgh]{A.A.~Koulbardis}
 \author[Petersburgh]{S.P.~Kruglov}
 \author[Basel]{B.~Krusche}
 \author[Moscow]{V.~Lisin}
 \author[Glasgow]{K.~Livingston}
 \author[Glasgow]{I.J.D.~MacGregor}
 \author[Basel]{Y.~Maghrbi}
 \author[Kent]{D.M.~Manley}
 \author[Washington]{Z.~Marinides}
 \author[Indonesia]{T. Mart}
 \author[Mainz]{M.~Martinez}
 \author[Glasgow]{J.C.~McGeorge}
 \author[Glasgow]{E.F.~McNicoll}
 \author[Sackville]{D.G.~Middleton}
 \author[Pavia]{A.~Mushkarenkov}
 \author[UCLA]{B.M.K.~Nefkens}
 \author[HISK]{A.~Nikolaev}
 \author[Petersburgh]{V.A. Nikonov}
 \author[Basel]{M.~Oberle}
 \author[Mainz]{M.~Ostrick}
 \author[Mainz]{P.B.~Otte}
 \author[Mainz, Washington]{B.~Oussena}
 \author[Pavia]{P.~Pedroni}
 \author[Basel]{F.~Pheron}
 \author[Moscow]{A.~Polonski}
 \author[UCLA]{S.~Prakhov}
 \author[Glasgow]{J.~Robinson}
 \author[Glasgow]{G.~Rosner}
 \author[Pavia]{T.~Rostomyan}
 \author[HISKP]{A.V. Sarantsev}
 \author[Mainz]{S.~Schumann}
 \author[Edinburgh]{M.H.~Sikora}
 \author[Catholic]{D.I.~Sober}
 \author[UCLA]{A.~Starostin}
 \author[Washington]{I. Strakovsky}
 \author[UCLA]{I.M.~Suarez}
 \author[Zagreb]{I.~Supek}
 \author[Giessen]{M.~Thiel}
 \author[HISKP]{A.~Thomas}
 \author[Mainz]{M.~Unverzagt}
 \author[Basel]{D.~Werthm\"uller}
 \author[Basel]{L.~Witthauer}
 \author[Basel]{F.~Zehr}
\address[Edinburgh]{School of Physics, University of Edinburgh, Edinburgh EH9 3JZ, UK}
\address[Mainz]{Institut f\"ur Kernphysik, University of Mainz, D-55099 Mainz, Germany}
\address[Glasgow]{Department of Physics and Astronomy, University of Glasgow, Glasgow G12 8QQ, UK}
\address[Kent]{Kent State University, Kent, Ohio 44242, USA}
\address[HISKP]{Helmholtz-Institut f\"ur Strahlen- und Kernphysik, University of Bonn, D-53115 Bonn, Germany}
\address[Petersburgh]{Petersburg Nuclear Physics Institute, 188300 Gatchina, Russia}
\address[Giessen]{Physikalisches Institut, University of Giessen, D-35392 Giessen, Germany}
\address[Pavia]{INFN Sezione di Pavia, I-27100 Pavia, Italy}
\address[Washington]{The George Washington University, Washington, DC 20052, USA}
\address[UCLA]{The George Washington University, Washington, DC 20052, USA}
\address[Lebedev]{Lebedev Physical Institute, 119991 Moscow, Russia}
\address[Basel]{Institut f\"ur Physik, University of Basel, CH-4056 Basel, Switzerland}
\address[Sackville]{Mount Allison University, Sackville, New Brunswick E4L3B5, Canada}
\address[Moscow]{Institute for Nuclear Research, 125047 Moscow, Russia}
\address[Zagreb]{Rudjer Boskovic Institute, HR-10000 Zagreb, Croatia}
\address[Indonesia]{Departemen Fisika, FMIPA, Universitas Indonesia, Depok 16424, Indonesia}
\address[Catholic]{The Catholic University of America, Washington, DC 20064, USA}
\fntext[label1]{jude@physik.uni-bonn.de, present address: Physikalisches Institut, Universit{\"a}t Bonn, Germany}
\fntext[label2]{Present address: Department of Physics and Astronomy, University of Glasgow, Glasgow, UK}
\fntext[label3]{dwatts1@ph.ed.ac.uk}

\begin{abstract}
Measurements of $\gamma p \rightarrow K^{+} \Lambda$ and $\gamma p \rightarrow K^{+} \Sigma^0$ cross-sections have been obtained with the
photon tagging facility and the Crystal Ball calorimeter at MAMI-C.  The measurement uses a novel $K^+$ meson identification technique in which the weak decay products are characterized using the energy and timing characteristics of the energy deposit in the calorimeter, a method that has the potential to be applied at many other facilities.
The fine center-of-mass energy ($W$) resolution and statistical accuracy of the new data results in a significant impact on partial wave analyses aiming to better establish the excitation spectrum of the nucleon. The new analyses disfavor a strong role for quark-diquark dynamics in the nucleon. 
\end{abstract}


\end{frontmatter}


\section{Introduction}\label{sec:intro}
Establishing the excitation spectrum of a composite system has historically been one of the most effective 
ways to determine the detailed nature of the interactions between its constituents. Establishing the excitation 
spectrum of the nucleon; a complex bound system of valence quarks, sea quarks and gluons,  is currently one 
of the highest priority goals of hadron and nuclear physics. The spectrum is a fundamental constraint on our
 understanding of the nature of QCD confinement in light quark systems. Recent advances in theory have linked 
the excitation spectrum to QCD via lattice predictions~\cite{durr08} and holographic dual theories~\cite{brodsky12}. 
These complement the phenomenological QCD-based models such as constituent quark models~\cite{capstick} and 
soliton models~\cite{diakonov}. 

Despite its importance, the spectrum of nucleon resonances remains poorly established with the basic properties 
(electromagnetic couplings, masses, widths) and even the existence of many excited states uncertain (for a review see Ref.~\cite{klempt}). 
In an attempt to address this shortcoming, real photon beams have been used to excite nucleon targets, providing 
accurate data to constrain partial-wave analyses (PWA) and reaction models used to extract information on the excitation 
spectrum~\cite{mart00,lee01,kaonmaidwebpage,saidwebpage,anisovich11,anisovich12}.  This is the choice method for such studies, as the 
photon probe has a well-understood interaction (QED) and polarization degrees of freedom (linear and circular). 
A major program of measurements utilizing polarized photon beams, polarized targets and final-state nucleon 
polarimeters is currently underway with the goal to achieve a ``complete'', model-independent measurement of 
photoproduction reactions.

 The process $\gamma p \rightarrow K^+ \Lambda$ has the lowest energy threshold for photoproduction reactions 
with final-state particles containing strange valence quarks. This is a crucial channel as many models predict 
that some poorly established or ``missing'' resonances couple strongly to strange decay channels~\cite{capstick98}. 
Isospin conservation demands that only $N^{*}$ and not $\Delta$ resonances contribute to the reaction,  simplifying 
the interpretation of the data.  The weak decay of the $\Lambda$ allows access to its polarization from the 
distribution of its decay particles and ensures that $\gamma p \rightarrow K^+ \Lambda$ will be the first photoproduction 
reaction measured with a complete set of experimental observables, providing a benchmark channel for PWAs.

Recent measurements of $\gamma p \rightarrow K^+ \Lambda$ have been obtained with the SAPHIR~\cite{B1_glander04,B1_tran98} and CLAS detectors~\cite{B1_bradford06, B1_mccracken10}. Unfortunately 
the cross-section data have discrepancies that lead to significant differences in the PWA solutions when using either 
data set (see Ref.~\cite{mart10} for a review), however, measurements of $\gamma p \rightarrow K^+ \Sigma^0$ with similar analysis procedures give closer agreement.

$\gamma p \rightarrow K^+ \Lambda$ data with fine center-of-mass energy ($W$) resolution would be an important constraint on the existence of narrow $N^{*}$ states~\cite{arndt,mart11}. 
A number of recent searches for narrow $N^{*}$ near 1700 MeV (Ref.~\cite{mart13} for example) were motivated by the prediction of a non-strange, 
nucleon-like, member of the anti-decuplet with strong photocoupling to the neutron~\cite{diakonov2}. 
In response to recent evidence, a speculative new $N^{*}$ state at 1685 MeV was included in the recent Particle Data Group listings~\cite{pdg12}. However, alternative explanations for the narrow structures are also offered based on interference structures arising from known resonances~\cite{shklyar07} or coupled-channel effects~\cite{doring}. Disentangling the cause of the narrow structure in this mass region is likely to require accurate cross-section and polarisation observables for a range of reaction channels.

\section{The experiment}\label{sec:exp}

The data presented here were taken with the Crystal Ball detector~\cite{starostin01} 
at the Mainz Microtron accelerator facility (MAMI-C)~\cite{kaiser08} in a beamtime of 430 hours.  The energy 
tagged photon beam of $\sim 10^{5}\gamma$~MeV$^{-1}$~sec$^{-1}$ was produced by impinging the MAMI-C 1557.4~MeV 
electron beam on a thin copper radiator, with the photon energy ($E_\gamma$) determined by momentum analysis of the recoil 
bremsstrahlung electrons in the Glasgow Photon Tagger~\cite{mcgeorge08}.  Photon energy resolutions in the range of 3~-~4~MeV were achieved, 
corresponding to resolutions in the center of mass energy, $W$, in the range of 1.0~-~2.4~MeV. 
The photon beam was incident on a 10~cm long liquid hydrogen target comprising 4.2 $\times$ 10$^{23}$ protons per cm$^2$. 

\begin{figure}
 \begin{center}
 \includegraphics[clip=true, width=4cm]{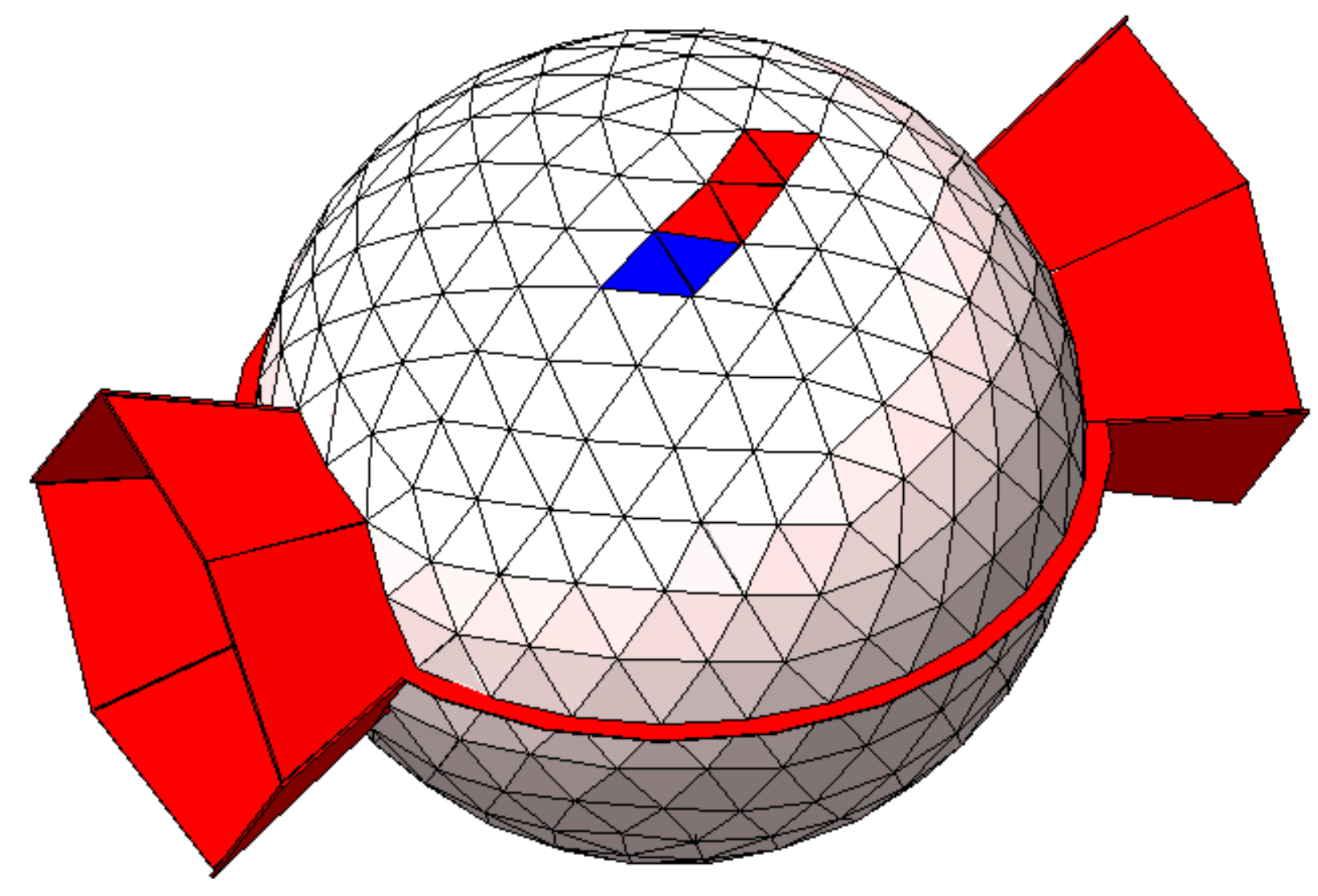}
 \end{center}
\caption{\label{fig:SimPic}The Crystal Ball in the Geant4 simulation. 
The shaded crystals have energy depositions following an incident $K^+$. The blue and red show the incident cluster and decay cluster respectively.}
\end{figure}

The Crystal Ball (Fig.~\ref{fig:SimPic}) is a segmented calorimeter of 672 NaI 
crystals covering 94\% of 4$\pi$ steradians.  Each crystal has separate TDC and ADC readouts giving a time 
resolution of 2-3~ns and a fractional energy resolution of $(1.7/E_\gamma)^{0.4}$~GeV.  
A Particle Identification Detector (PID), consisting of 24 plastic 
scintillators forming a cylinder~\cite{pid}, surrounded the target and gave an energy signal for charged particles. The experimental trigger 
required a total energy deposit in the Crystal Ball crystals of 360~MeV and at least 
three of 45 geometric trigger sections to fire. 

\section{$K^+$ identification in the Crystal Ball}\label{sec:kaonid}

\begin{figure}[t]
\begin{center}
\includegraphics[clip=true, width=9.0cm]{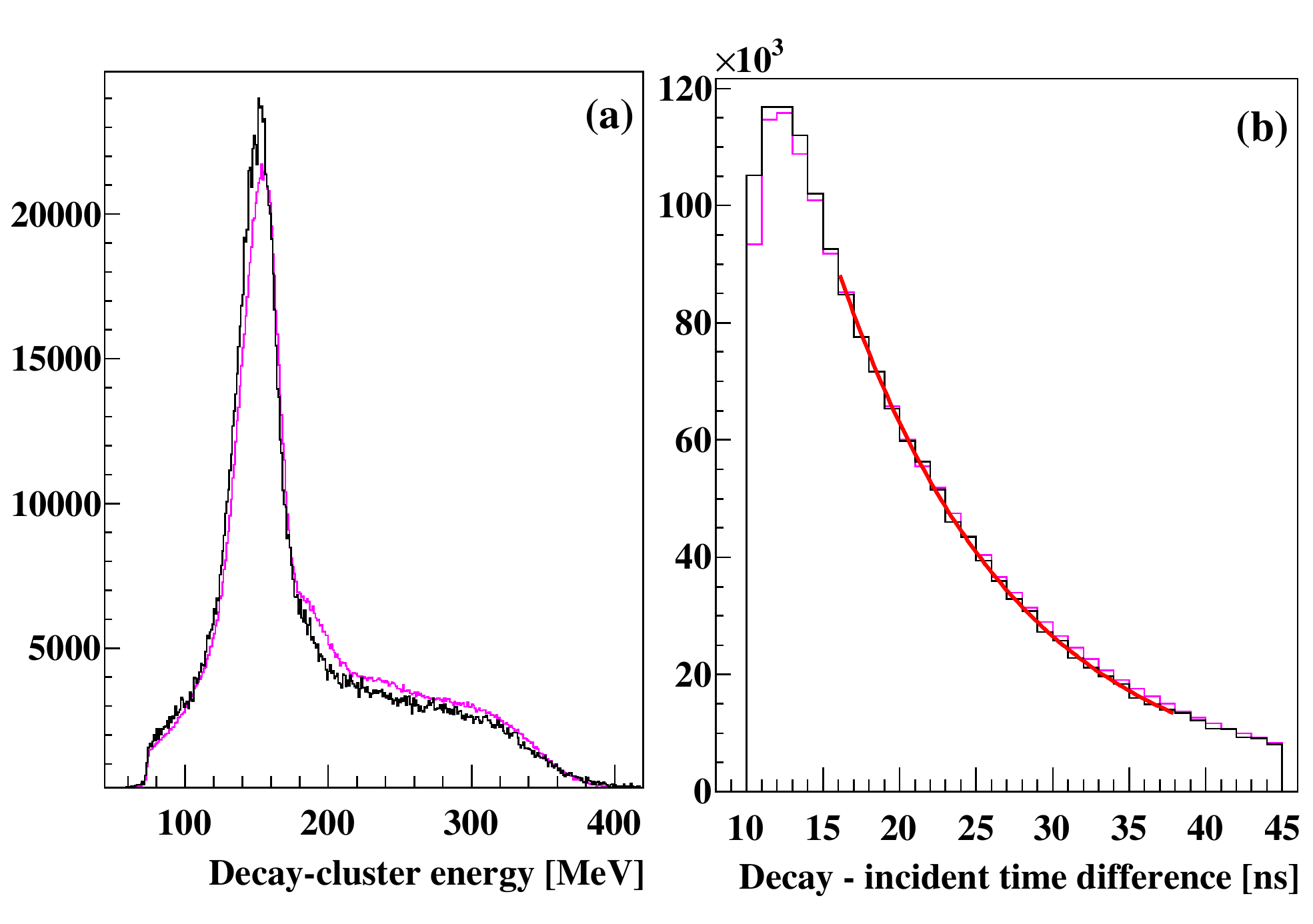}
\end{center}
\caption{\label{fig:ESideTSide} Decay-cluster characteristics for experimental and simulated 
data (black and magenta respectively): (a) Decay cluster summed energy. (b) Time difference between the incident cluster and decay cluster.}
\end{figure}

\begin{figure}[t]
\begin{center}
 \includegraphics[clip=true,width=9.0cm]{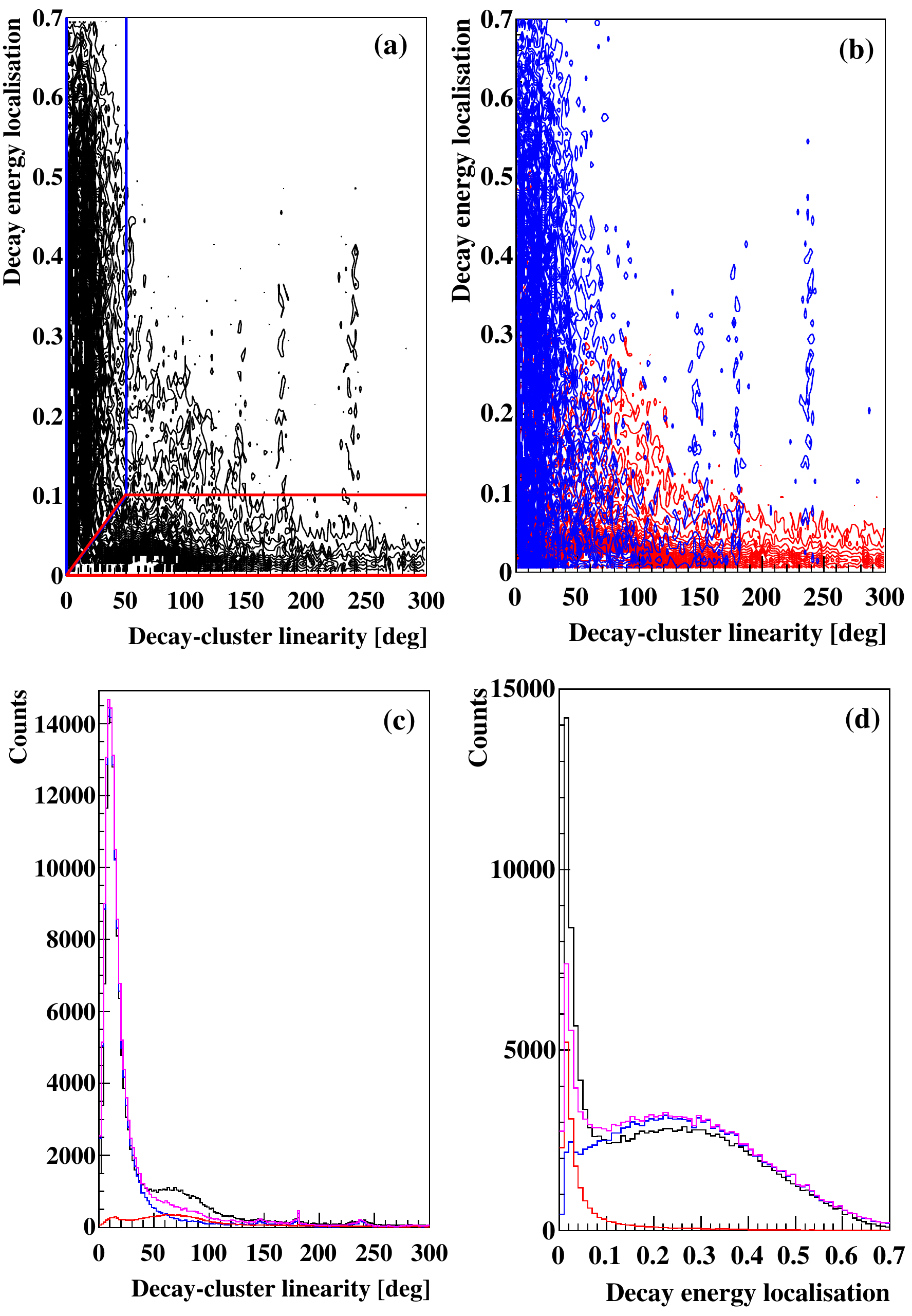}
\end{center}
 \caption{\label{fig:SplitDecayModes} Decay cluster parameters to select the $K^+$ decay mode.  (a)~Experimental data, with blue (red) selection cuts
to select muonic (pionic) decay modes.  (b)~Simulated data, with the muonic (pionic) mode shown in blue (red).  (c)~ The decay cluster linearity for
experimental data (black) and simulated data for muonic (pionic) in blue (red). (d)~The decay energy localisation  for
experimental data (black) and simulated data for muonic (pionic) in blue (red).  The simulated data has been scaled to the integral of
 the experimental data.} 
 \end{figure}

The extraction of $K^+\Lambda$ and $K^+\Sigma^0$ channels is complicated by the much larger yields from non-strange channels.  This work pioneers a new method of identifying $K^+$ in which its weak decay products are characterized by using the energy and timing characteristics of the detector hits in a segmented calorimeter.  
The two dominant decay modes are $K^+ \rightarrow \mu^+\nu_\mu$ (muonic) and $K^+ \rightarrow\pi^+\pi^0$ (pionic), with branching ratios of 64\% and 21\% respectively. The validity of the new technique was tested extensively 
by comparing a full Geant4~\cite{geant} simulation of the apparatus with the experimental data.  The main 
results of these studies are presented in Figs.~\ref{fig:ESideTSide} and \ref{fig:SplitDecayModes} and discussed below. 

Each cluster of hit crystals produced from a charged particle 
event in the Crystal Ball was separated into two ``sub-clusters''. The ``incident-cluster'' (IC) comprised those crystals 
having a timing coincidence within $\pm 3\sigma$ of the timing of the photoreaction in the target, where $\sigma$ 
is the achievable coincidence timing resolution ($\sim$~3~ns). Only events with a summed IC energy above 25 MeV 
and consisting of only one or two crystals were retained. The crystals with coincidence times at least 10~ns later than the photoreaction 
were assumed part of the ``decay-cluster'' (DC) from the decay of the stopped $K^+$. A minimum summed DC energy 
of 75 MeV and at least 4 crystals in the DC was required. A cluster pattern for a typical muonic decay event visualized in 
the Geant4 simulation is presented in Fig.~\ref{fig:SimPic}. 

Fig.~\ref{fig:ESideTSide} shows the energy spectrum for the DC, exhibiting a peak at 150~MeV 
consistent with the energy of the $\mu^+$ from $K^+ \rightarrow \mu^+\nu_\mu$ decay at rest.  A shoulder
 extends to 350~MeV, which is the maximum energy deposition for the pionic decay ($K^+ \rightarrow\pi^+\pi^0$).  Fig.~\ref{fig:ESideTSide}(b) is the 
time difference between the IC and DC. An exponential fit gives a lifetime in agreement with the accepted $K^+$ lifetime of approximately 12~ns.

To separate events into the two dominant $K^+$ decay modes, two parameters were used:  The fractional energy in
the furthest crystal in the DC with respect to the total energy in the DC (the decay energy localisation), and the average difference in angle between each crystal in the DC and the IC
(the decay cluster linearity).  Fig.~\ref{fig:SplitDecayModes}  shows these parameters plotted for both experimental and simulated data. Good agreement between data and simulation is observed, with small deviations only evident in regions where the pionic decay mode dominates. The pionic decays have an increased sensitivity to the systematics of the modelling of the low energy thresholds in the CB crystals in the simulation. For this reason, and also because the shower shape gave an improved $K^+$ momentum reconstruction, only the dominant muonic decay events were retained for further analysis by applying the two dimensional selection cuts in Fig.~\ref{fig:SplitDecayModes}(a). A small proportion of pionic decay events and other decay modes (such as  $K^+ \rightarrow\pi^0 e^+ \nu_e$ with a branching ratio of 5\%) were expected to remain in the yield.  The IC summed energy was then utilised in a $\Delta E-E$ analysis with the $\Delta E$ provided by the signal in the PID and used to reconstruct the momentum of the $K^{+}$. 

This new $K^+$ identification technique enables $K^+$ detection without the need for large scale spectrometers or Cerenkov detectors and has wide applicability to other facilities. The technique has already been incorporated into the BGO-OD experiment~\cite{hartmut} and will be the basis of a new online $K^+$ trigger at MAMI, significantly increasing future $K^+$ yields.  The technique is also a viable method for identifying $K^+$ in fast timing environments such as in laser plasma based accelerators.

\section{Extracting $K^+\Lambda$ and $K^+\Sigma^0$ differential cross-sections}\label{sec:splitchannels}

\begin{figure}[]
\begin{center}
\includegraphics[clip=true, width=9.5cm]{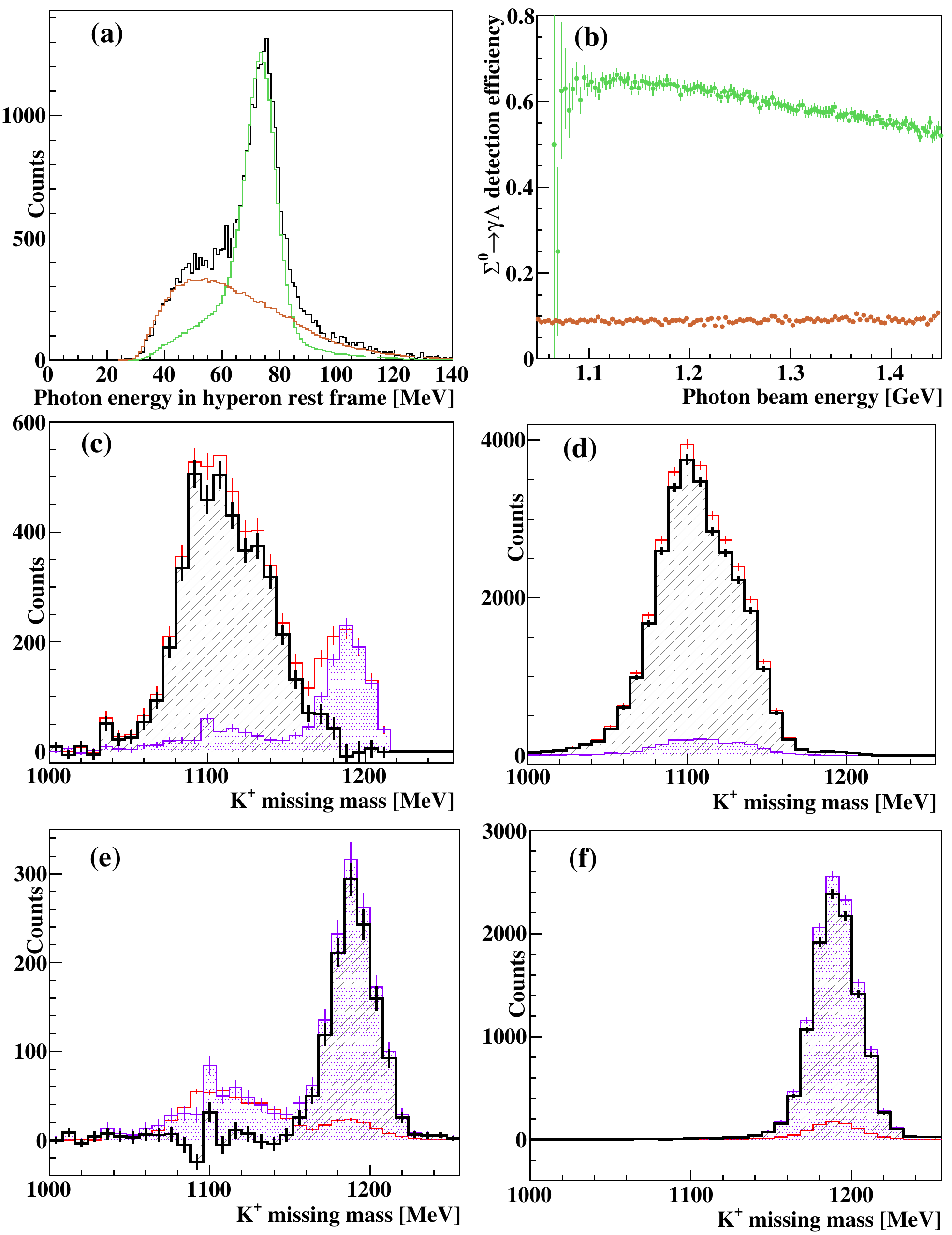}
\end{center}
\caption{\label{fig:ESideTSide2}
(a)~Neutral particle energies in the hyperon rest frame for experimental data (black), simulated $K^+\Lambda$ data (orange) and simulated $K^+\Sigma^0$ data (green).
(b)~Simulated decay-$\gamma$ detection efficiency for $\Sigma^0\rightarrow \Lambda\gamma$ (green), and false decay-$\gamma$ detection efficiency for $K^+ \Lambda$ (orange).
(c)~Experimental data for the missing mass recoiling from the $K^+$ for the interval $1.086 < E_\gamma < 1.229$~GeV and $\cos_K^{CM} = -0.1$,  used to extract the $K^+\Lambda$ yield.  Without (with and efficiency corrected)
a decay-$\gamma$ candidate shown by unfilled red (filled violet) lines, and the subtracted $K^+\Lambda$ yield (thick black line, shaded fill).
(d)~Simulated $K^+ \Lambda$ yield for the same scenario as (c).
(e)~Experimental data for the missing mass recoiling from the $K^+$ for the same interval as in (c), used to extract the $K^+\Sigma^0$ yield.  Without, and efficiency corrected (with)
a decay-$\gamma$ candidate shown by unfilled red (filled violet) lines, and the subtracted $K^+\Sigma^0$ yield (thick black line, shaded fill).
(f)~Simulated $K^+ \Sigma^0$ yield for the same scenario as (e).
}
\end{figure}

 A new technique to cleanly separate $\gamma p \rightarrow K^+\Lambda$ and $K^+\Sigma^0$ yields was used, via the identification of the decay $\Sigma^0\rightarrow \Lambda \gamma$ in the Crystal Ball.  
Fig.~\ref{fig:ESideTSide2}(a) shows the energy of neutral particles detected in coincidence with the $K^{+}$, boosted into the rest frame of the hyperon. The peak at 77~MeV is from the detection of the $\gamma$ from the $\Sigma^0$ decay,  
having an energy corresponding to the $\Sigma^0$-$\Lambda$ mass difference.  Events with energies between 55-95~MeV were selected as decay-$\gamma$ candidate events for $\Sigma^0\rightarrow \Lambda \gamma$. 
From Fig.~\ref{fig:ESideTSide2}(a) it is clear a background of additional uncharged events is also present, arising from photons or neutrons from $\Lambda$ decays.
The decay-$\gamma$ detection efficiency was determined with simulated data to behave linearly with $E_\gamma$ and to be approximately 60\%.  The false  decay-$\gamma$ detection efficiency from $K^+\Lambda$ events was approximately 9\% (Fig.~\ref{fig:ESideTSide2}(b)).

Fig.~\ref{fig:ESideTSide2}(c) shows the reconstructed missing-mass of the system recoiling from the $K^{+}$ over a restricted kinematic range.  The $\Sigma^0$ and $\Lambda$ masses are clearly visible, with the relative contribution of the $\Sigma^0$ enhanced with the requirement of a decay-$\gamma$ candidate.  The yield of coincident decay-$\gamma$ events (filled violet line) has been scaled according to the decay-$\gamma$ detection efficiency.  
This efficiency corrected yield was used to subtract the $\Sigma^{0}$ contribution from the data for each kinematic bin. The remaining yield attributed to $K^{+}\Lambda$ after this subtraction is shown by the 
filled shaded thick black line in Fig.~\ref{fig:ESideTSide2}(c).

The yield of simulated $K^{+}\Lambda$ events for the same kinematic bin are shown in Fig.~\ref{fig:ESideTSide2}(d). The simulation reproduces the shape of the $K^{+}\Lambda$ distribution observed in the experiment. The relative contribution of misidentified decay-$\gamma$ events to the yield, arising from the $\Lambda$ decay products, is shown by the filled violet line under the $\Lambda$ mass peak. These misidentified events reduce the extracted $K^{+}\Lambda$ yield by approximately 5\%.  This loss in yield however cancels out in the cross-section calculation as the same analysis procedure is applied to the simulated data which is used to determine the detection efficiency.

Fig.~\ref{fig:ESideTSide2}(e,f) show the same experimental and simulated missing mass data, however the yield of non-coincident decay-$\gamma$ events have been scaled according to the 
false decay-$\gamma$ detection efficiency from $K^+\Lambda$ events (thin solid red line).
This corrected yield was used to subtract $K^+\Lambda$ contributions to leave only contributions attributed to $K^+\Sigma^0$ events (filled shaded thick black lines).  

Fitting Gaussian functions to the total missing-mass spectra to separate the reaction channels gave an agreement with the above method to better than 4\%, which is taken as the estimated systematic error.

Detection efficiencies were obtained by analyzing Geant4-simulated $K^+\Lambda$ and $K^+\Sigma^0$ events including appropriate timing and 
energy resolutions and using angular distributions from the SAID PWA~\cite{saidwebpage}.  Experimental trigger conditions were implemented as described in Ref.~\cite{sergey}.  A maximum detection efficiency 
of approximately 10\% was achieved.

The modelling of $K^+$ hadronic interactions in the Crystal Ball gave a maximum systematic uncertainty of 4\% to the yield, increasing with $E_\gamma$.  This was assessed from comparisons of different physics models in the simulation, and by switching off hadronic interactions.  Contamination in the $K^+\Lambda$ yield from other channels
passing the selection cuts (dominantly $\gamma p \rightarrow p \pi^+ \pi^-$) gave an uncertainty typically less than 4\% and only at very backward angles.  The required identification of the decay-$\gamma$ for $K^+\Sigma^0$ rendered contamination in the $K^+\Sigma^0$ yield from other channels negligible.  Systematic effects from the modelling of the experimental trigger in the simulation  (estimated from a 10~MeV variation of the Crystal Ball energy sum threshold) were typically 4\% near threshold for $K^+\Lambda$ and reducing with $E_\gamma$.  For $K^+\Sigma^0$, the uncertainty was 2-3\% near threshold and only at very backward angles.  Systematic errors from non-hydrogen components of the target cell, target cell length and PID efficiency were each less than 1\%.  

\section{Results and interpretation}\label{sec:results}

\begin{figure}
\begin{center}
\includegraphics[trim=0cm 0cm 0cm 0cm, clip=true, width=0.52\textwidth]{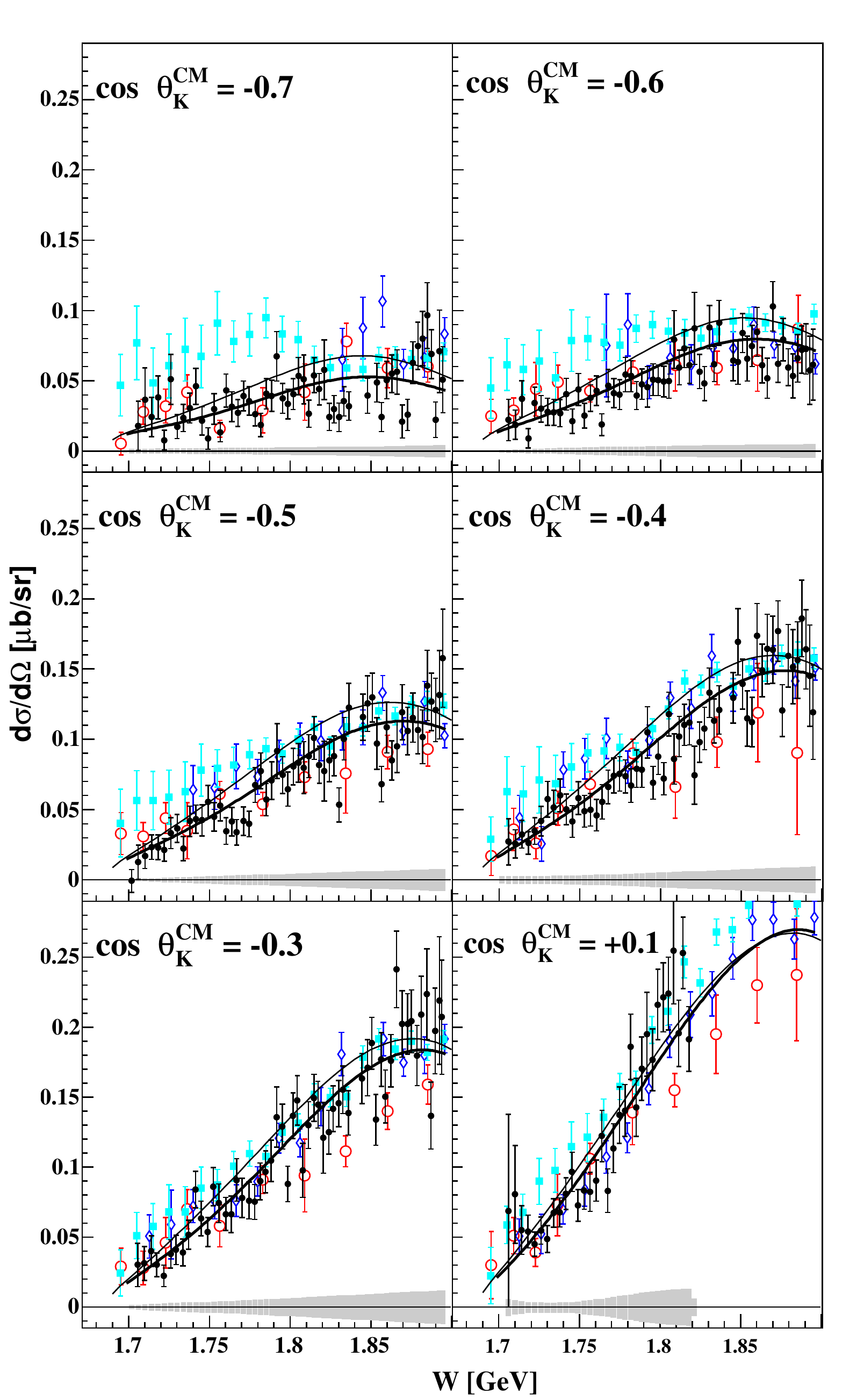}
\end{center}
\caption{\label{fig:DiffCrossSectionsKSigma}Differential $\gamma p \rightarrow K^+ \Sigma^0$ cross-sections versus $W$. 
Black filled circles are the current data with systematic uncertainties plotted gray on the abscissa.  Red open circles are SAPHIR data~\cite{B1_glander04}, blue open diamonds are CLAS data of Bradford \textit{et al.}~\cite{B1_bradford06}, cyan solid squares are CLAS data of Dey \textit{et al.}~\cite{Dey10}.  The thin black line is the current BnGa 2011-2 solution~\cite{anisovich12} and the thick black line is the  BnGa 2011-02 solution including the new $K^+\Lambda$ and $K^+\Sigma^0$ data~\cite{anisovich}.
(The SAPHIR data have $\cos \theta_{K}^{CM}$ intervals backwards by 0.05 than the given values.)}
\end{figure}

\begin{figure}
\begin{center}
\includegraphics[trim=0cm 0cm 0cm 0cm, clip=true, width=0.52\textwidth]{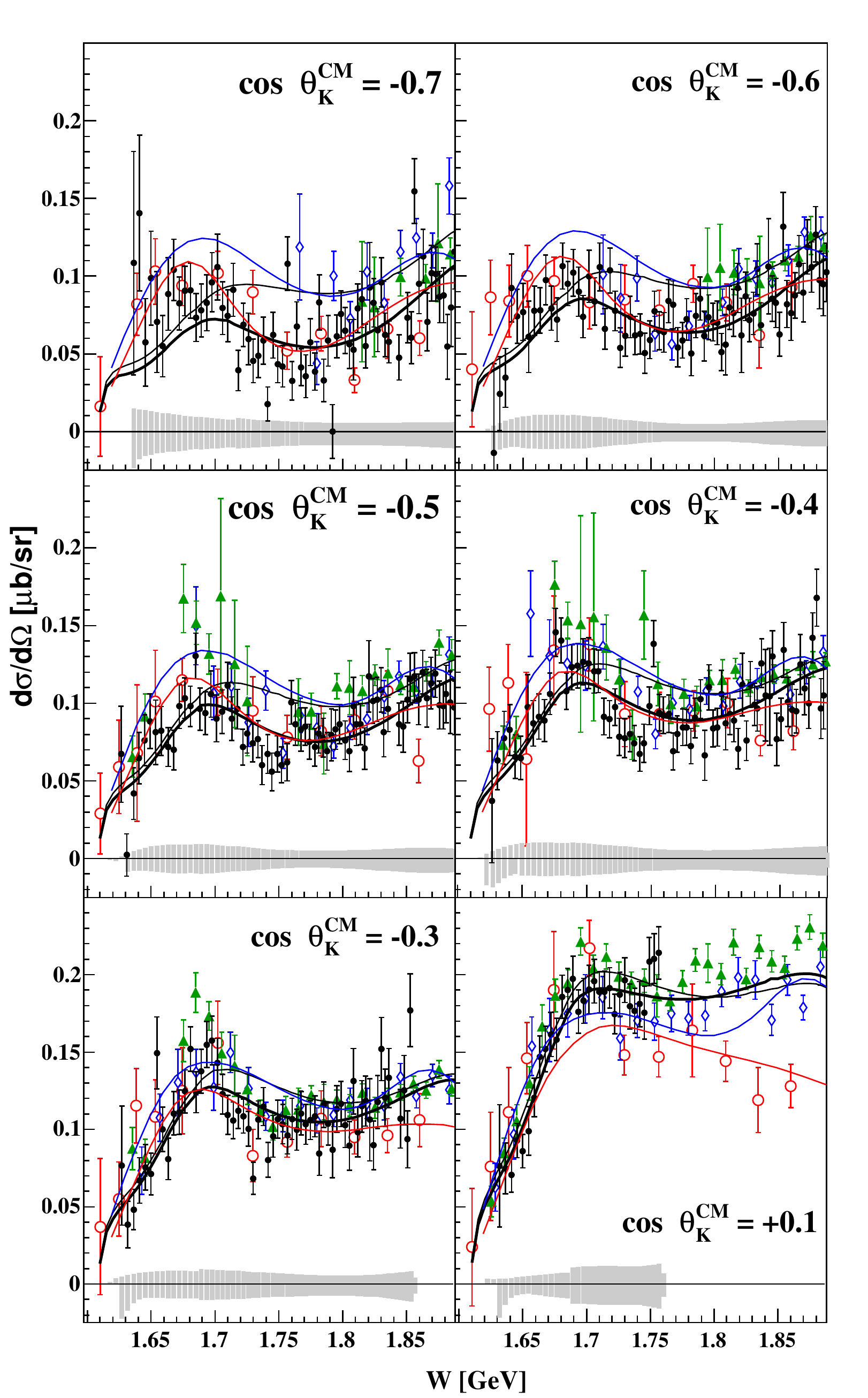}
\end{center}
\caption{\label{fig:DiffCrossSections}Differential $\gamma p \rightarrow K^+ \Lambda$ cross-sections versus $W$.
Black filled circles are the current data with systematic uncertainties plotted gray on the abscissa.  
Red open circles are SAPHIR data~\cite{B1_glander04}, 
blue open diamonds are CLAS data of Bradford \textit{et al.}~\cite{B1_bradford06} and 
green filled triangles are CLAS data of McCracken \textit{et al.}~\cite{B1_mccracken10}.  
The thin black line is the current BnGa 2011-2 solution~\cite{anisovich12} and the thick black line is the  BnGa 2011-02 solution including the new $K^+\Lambda$ and $K^+\Sigma^0$ data~\cite{anisovich}.
The thin red and blue lines are fits from the KM model to SAPHIR data and CLAS data~\cite{mart12} respectively. 
(The SAPHIR data have $\cos \theta_{K}^{CM}$ intervals backwards by 0.05 than the given values.)
}
\end{figure}

\begin{figure}
\begin{center}
\includegraphics[trim=0cm 0cm 0cm 0cm, clip=true, width=6cm]{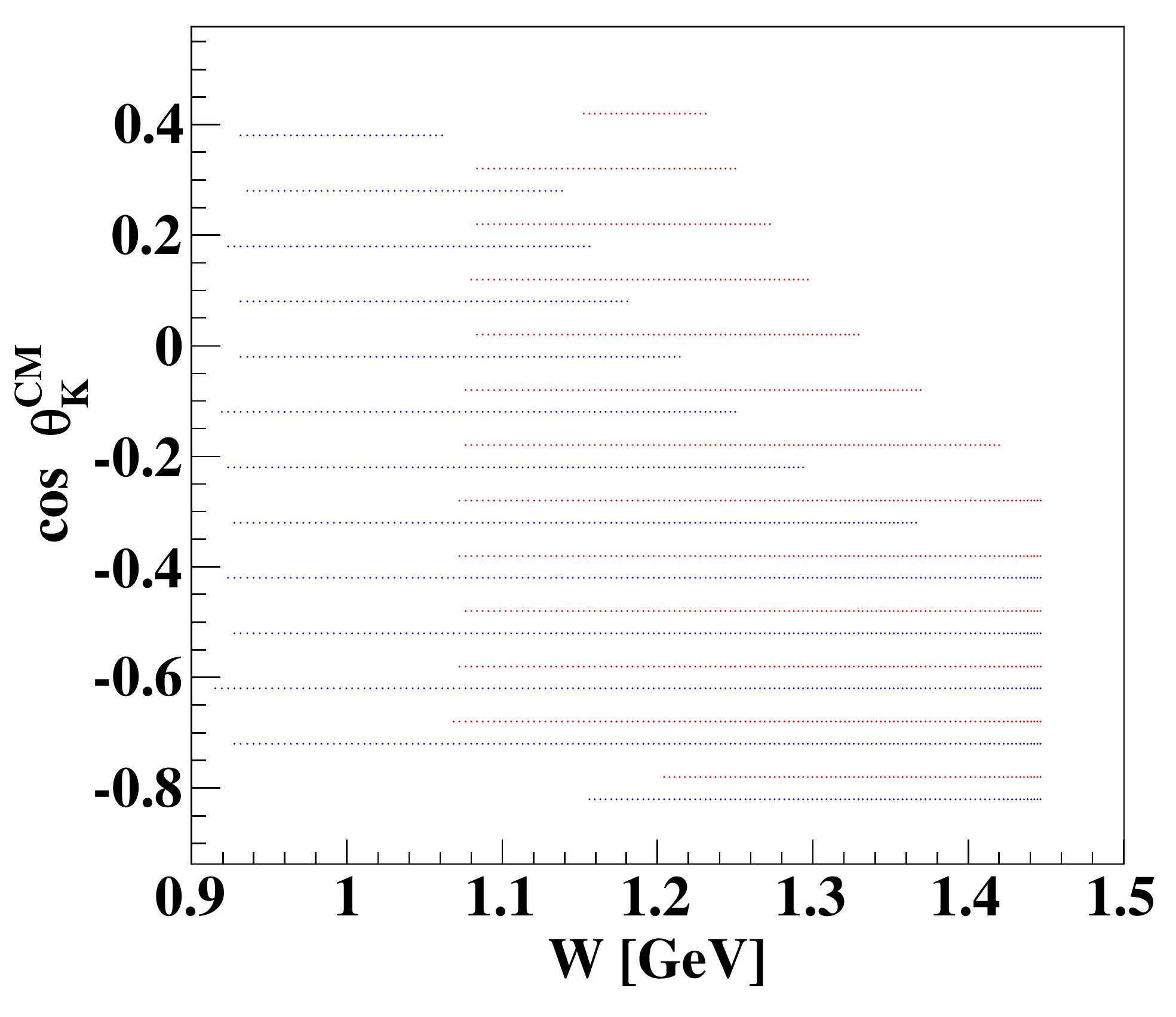}
\end{center}
\caption{\label{fig:KinematicRange} Kinematic range of differential cross-section measurements for $K^+\Lambda$ (blue) and $K^+\Sigma^0$ (red).  data points are offset in angle to avoid an overlap.}
\end{figure}

The quality of the new Crystal Ball data is illustrated in Figs.~\ref{fig:DiffCrossSectionsKSigma} and \ref{fig:DiffCrossSections}, where cross-sections for $K^+\Sigma^0$ and $K^+\Lambda$ as a function of $W$ are 
shown for selected center-of-mass $K^+$ polar angle bins ($\theta_{K}^{CM}$), compared with previous SAPHIR~\cite{B1_glander04} and CLAS~\cite{B1_bradford06, Dey10, B1_mccracken10} 
data.   For clarity, the data are rebinned by a factor of two, however the attainable $W$ resolution of the new data is a factor of 4 to 10 improvement over previous data. After normalising for the different widths in binning, the statistical accuracy of the new data is typically a factor of 1.5 better than previous data, except at forward $K^+$ angles where the accuracy is comparable.  The kinematic range the new data cover is shown in Fig.~\ref{fig:KinematicRange}.

The new $\gamma p \rightarrow K^+ \Sigma^0$ data (Fig.~\ref{fig:DiffCrossSectionsKSigma}) are consistent with the world data over most of the angular range, demonstrating that systematic errors for the new detection techniques are well understood. At backward angles there is divergence between the previous measurements and the new data give better agreement with the SAPHIR data.~\cite{B1_glander04}. The  $\gamma p \rightarrow K^+ \Lambda$ data show general agreement with the previous data and confirm the existence of the broad peak like structure centered around 1670~MeV for backward $K^+$ angles.
A demonstration of the compatability between this data and previous CLAS data is apparent through the fitting 
of the Bonn-Gatchina PWA~\cite{anisovich11} BG2011-02 solution (a description is given below).  From threshold to $W=1.9$~GeV, the solution describes the CLAS data with a $\chi^2$ of 0.57 for $K^+\Lambda$~\cite{B1_mccracken10} and 1.43 for $K^+\Sigma^0$~\cite{Dey10}.  
With the additional constraints of this new data,
the solution describes the the CLAS data with a $\chi^2$
of 0.54 for $K^+\Lambda$ and 1.90 for $K^+\Sigma^0$~\cite{andrey}.
The significant increase in $\chi^2$ for the $K^+\Sigma^0$ data is mostly attributed to the discrepancy at backward angles close to threshold. 

The data are compared to predictions from PWAs in the Kaon Maid (KM)~\cite{kaonmaidwebpage} and 
Bonn-Gatchina (BnGa)~\cite{anisovich11} framework, which are constrained by the various combinations
 of data sets indicated in the figure captions.
The main quoted systematic error in extracting resonance properties in the BnGa PWA framework is the existence of two solutions (BG2011-01 and BG2011-02) which give a similarly low $\chi^{2}$  when fitted to the world database. The solutions have different resonance contributions and helicity couplings (for a detailed description see \cite{anisovich11}). The addition of the new $K^+\Sigma^0$ and $K^+\Lambda$ data resolve these solutions. Only BG2011-02 can fit the new data and the world dataset with a satisfactorily low $\chi^{2}$ of 1.3 and 1.2 for $K^+\Sigma^0$ and $K^+\Lambda$ respectively~\cite{andrey}.  The most significant difference between the solutions is that BG2011-2 supports the need for two $P_{13}$ nucleon resonances close in mass: a $P_{13}$(1900) and $P_{13}$(1975). Despite constituent quark models (CQM) predicting the existence of two $3/2^+$ nucleon states in the region 1850-2000 MeV, it is 
difficult to explain two such states in the framework of a quark-diquark model or under the assumption of chiral symmetry restoration. The new experimental data therefore provide new constraints on the dynamics of quarks within the nucleon~\cite{anisovich11}.
  
The well defined structure in the $K^+\Lambda$ cross-section around 1670 MeV at backward $K^+$ angles provides a valuable constraint on the existence and width of the disputed~\cite{arndt1710} P$_{11}$(1710) resonance. To fit the structure a 30\% reduction in the resonance width was necesary in the BnGa analysis. Interestingly this produces a width now consistent with the other sightings of this resonance~\cite{klempt}.

The improved statistical accuracy and $W$ resolution of the new data allows constraints on the existence of structures in the cross-section arising from narrow resonance states or coupled channel effects.  There are indications of structure between 1650~-~1700~MeV and at 1740~MeV which are not described by any of the PWA models.

\begin{figure}
\begin{center}
\includegraphics[clip=true, width=6cm]{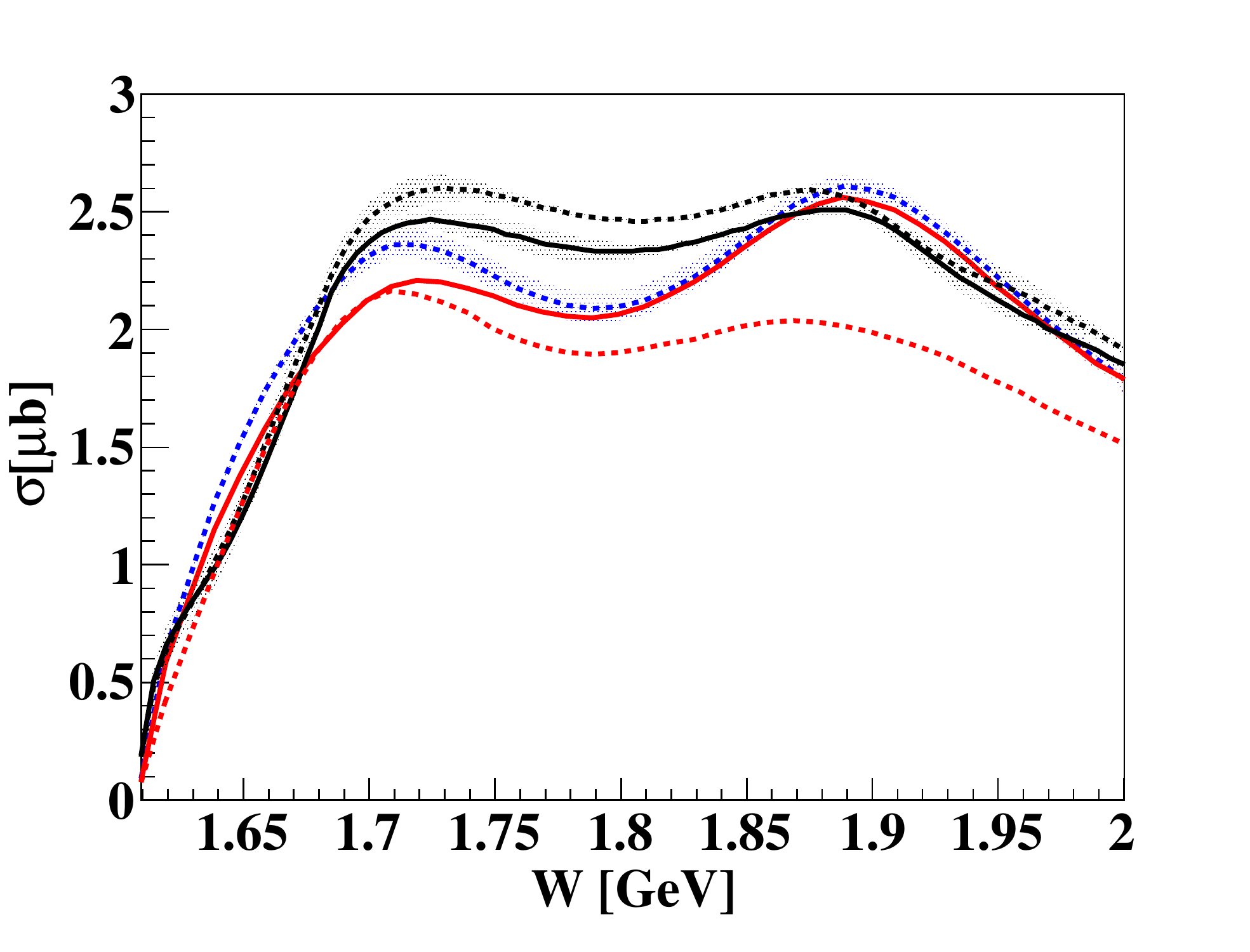}
\end{center}
\caption{\label{fig:TotCrossSections}Total $\gamma p \rightarrow K^+ \Lambda$ cross-section versus $W$.
The thin black dashed line is the current BnGa 2011-2 solution~\cite{anisovich12} and the thick solid black line is the  BnGa 2011-02 solution including the new $K^+\Lambda$ and $K^+\Sigma^0$ data~\cite{anisovich}.
The thin red dashed line is the current KM model constrained by SAPHIR data~\cite{mart12}, the thin blue dashed line is the current KM model constrained by CLAS data, and the thick red line is the KM model constrained by CLAS data and this current $K^+\Lambda$ data~\cite{martcommunication}.  Shaded bands estimate systematic errors.} 
\end{figure}

The total cross-section for $\gamma p \rightarrow K^+ \Lambda$ has been debated in the literature where structure around 1900~MeV has been largely interpreted as evidence for a missing $D_{13}$ resonance (for example~\cite{mart10} and references therein).  Constraints from the new data lead to revised extrapolated total cross-sections as shown in Fig.~\ref{fig:TotCrossSections}. The cross-sections extrapolated using the BnGa 2011-02 solution are reduced below 1900~MeV with the inclusion of the new data. The cross-sections extrapolated from the KM model show a reduction mainly in the region around the first peak in the cross-section at 1700~MeV. The structure at 1900~MeV is still evident in the revised extrapolations.  

\section{Conclusions}\label{sec:conclusions}

Precision measurements of the $\gamma p \rightarrow K^+ \Lambda$ and $\gamma p \rightarrow K^+ \Sigma^0$ differential cross-section have been obtained with a new $K^{+}$ identification method which has wide applicability for other facilities. 
 The new  $\gamma p \rightarrow K^+ \Lambda$ and $\gamma p \rightarrow K^+ \Sigma^0$ data have significantly improved center-of-mass energy resolution than previous data and provide a significant new constraint to the world database of meson photoproduction.  A combined analysis of both reaction channels resolves the largest quoted systematic error in determining the resonance spectrum with the BnGa PWA framework, resulting in a preference for a nucleon resonance spectrum which is inconsistent with models assuming quark-diquark dynamics in the nucleon.

The authors acknowledge the excellent support
of the MAMI-C accelerator group.  This work was supported by the UK STFC, the German DFG (SFB 443), SFB/Tr16, the Schweizerischer Nationalfonds and the European Community-Research 
Infrastructure Activity under the FP6 ``Structuring the European Research Area'' program (HadronPhysics, Contact No. RII3-CT-2004-506078), DFG-RFBR
(Grant No. 05-02-04014), the U.S DOE, U.S. NSF and NSERC (Canada).



\begin{thebibliography}{00}
\bibitem{durr08} R.G. Edwards et al., Phys. Rev. D 84 (2011) 074508.
\bibitem{brodsky12} S.J. Brodsky and G.F. de Teramond, Few-Body Syst. 52 (2012) 203. 
\bibitem{capstick} S. Capstick and W. Roberts, Prog. Part. Nucl. Phys. 45 (2000) s241.   
\bibitem{diakonov} D. Diakonov and V. Petrov, World Scientific, 1 (2000) 359, arxiv:hep-ph/0009006. 
\bibitem{klempt} E. Klempt and J.-M. Richard, Rev. Mod. Phys. 82 (2010) 1095. 
\bibitem{mart00} T. Mart, Phys. Rev. C 62 (2000) 038201.  
\bibitem{lee01} F.X. Lee, T. Mart, C. Bennhold, H. Haberzettl and L.E. Wright, Nucl. Phys. A 695 (2001) 237.  
\bibitem{kaonmaidwebpage}http://www.kph.uni-mainz.de/MAID/kaon/kaonmaid.html.  
\bibitem{saidwebpage} R.A. Arndt, W.J. Briscoe, I.I. Strakovsky and R.L. Workman. SAID  
Partial-Wave Analysis Facility, gwdac.phys.gwu.edu (2012).
\bibitem{anisovich11} A.V. Anisovich et al., Eur. Phys. J. A 47 (2011) 153. 
\bibitem{anisovich12} A.V. Anisovich et al., Eur. Phys. J. A 48 (2012) 15.  
\bibitem{capstick98} S. Capstick and W. Roberts, Phys. Rev. D 58 (1998) 074011.
\bibitem{B1_glander04} K.H. Glander et al., Eur. Phys. J. A 19 (2004) 251.
\bibitem{B1_tran98} M.Q. Tran et al., Phys. Lett. B 445 (1998) 20.
\bibitem{B1_bradford06} R. Bradford and R.A. Schumacher et al. (CLAS Collaboration), Phys. Rev. C 73 (2006) 035202.

\bibitem{Dey10} B. Dey, C.A. Meyer, M. Bellis, M.E. McCracken, M. Williams et al. (CLAS Collaboration), Phys. Rev. C 82 (2010) 025202. 

\bibitem{B1_mccracken10} M.E. McCracken, M. Bellis, C.A. Meyer, M. Williams et al., Phys. Rev. C 81 (2010) 025201.
\bibitem{mart10} T. Mart, Int. J. Mod. Phys. E 19 (2010) 2343. 
\bibitem{arndt} R.H Arndt, Ya.I. Azimov, M.V. Polyakov, I.I. Strakovsky and R.L. Workman, Phys. Rev. C 69 (2004) 035208. 
\bibitem{mart11} T. Mart, Phys. Rev. D 83 (2011) 094015. 
\bibitem{mart13} T. Mart, Phys. Rev. D 88 (2013) 057501. 
\bibitem{diakonov2} V. Petrov, and M. V. Polyakov, Z Phys A 359 (1997) 305. 
\bibitem{pdg12} J. Beringer et al. (Particle Data Group), Phys. Rev. D 86 (2012) 010001.
\bibitem{shklyar07} V. Shklyar, H. Lenske and U. Mosel, Phys. L$K^+\Sigma^0$ett. B 650 (2007) 172.
\bibitem{doring} M. Doring and K. Nakayama, Phys. Lett. B 683 (2010) 145; A.V. Anisovich et al., Phys. Lett. B 719 (2013) 89. 
\bibitem{starostin01} A. Starostin et al., Phys. Rev. C 64 (2001) 055205. 
\bibitem{kaiser08} K.-H. Kaiser et al., Nucl. Instr. Meth. A 593 (2008) 159. 
\bibitem{mcgeorge08} J.C. McGeorge et al., Eur. Phys. J. A 37 (2008) 129. 
\bibitem{pid} D.P. Watts, \textit{Proceedings of the 11th International Conference on
Calorimetry in Particle Physics}, Perugia, Italy, World Scientific, (2005) 560. 
\bibitem{hartmut}H. Schmieden, Int. J. Mod. Phys. E 19 (2010) 1043.
\bibitem{geant}J. Allison et al., IEEE Trans. on Nucl. Sci. 53 (2006) 270. 
\bibitem{sergey}E.F. McNicoll, S. Prakhov and I.I. Strakovsky et al., Phys. Rev. C 82 (2010) 035208. 
\bibitem{anisovich} A.V. Anisovich, private communication (2012). 
\bibitem{mart12} T. Mart and M.J. Kholili, Phys. Rev. C 86 (2012) 022201.  
\bibitem{andrey} A.V. Sarantsev and V. Nikonov, private communication (2012). 
\bibitem{arndt1710}R. A. Arndt,W. J. Briscoe, I. I. Strakovsky, and R. L.Workman, Phys. Rev. C 79 (2009) 065207.
\bibitem{martcommunication} T. Mart, private communication (2012). 
\bibitem{Martinez} A. Martinez Torres et al.,  Phys. Rev. C 74 (2006) 045205.

 \end{thebibliography}
\end{document}